\documentclass[12pt,preprint]{emulateapj}

\def\reff@jnl#1{{\rm#1\/}}

\def\aj{\reff@jnl{AJ}}                  
\def\araa{\reff@jnl{ARA\&A}}            
\def\apj{\reff@jnl{ApJ}}                
\def\apjl{\reff@jnl{ApJ}}               
\def\apjs{\reff@jnl{ApJS}}              
\def\ao{\reff@jnl{Appl.Optics}}         
\def\apss{\reff@jnl{Ap\&SS}}            
\def\aap{\reff@jnl{A\&A}}               
\def\aapr{\reff@jnl{A\&A~Rev.}}         
\def\aaps{\reff@jnl{A\&AS}}             
\def\azh{\reff@jnl{AZh}}                
\def\baas{\reff@jnl{BAAS}}              
\def\jrasc{\reff@jnl{JRASC}}            
\def\memras{\reff@jnl{MmRAS}}           
\def\mnras{\reff@jnl{MNRAS}}            
\def\pra{\reff@jnl{Phys.Rev.A}}         
\def\prb{\reff@jnl{Phys.Rev.B}}         
\def\prc{\reff@jnl{Phys.Rev.C}}         
\def\prd{\reff@jnl{Phys.Rev.D}}         
\def\prl{\reff@jnl{Phys.Rev.Lett}}      
\def\pasp{\reff@jnl{PASP}}              
\def\pasj{\reff@jnl{PASJ}}              
\def\qjras{\reff@jnl{QJRAS}}            
\def\skytel{\reff@jnl{S\&T}}            
\def\solphys{\reff@jnl{Solar~Phys.}}    
\def\sovast{\reff@jnl{Soviet~Ast.}}     
\def\ssr{\reff@jnl{Space~Sci.Rev.}}     
\def\zap{\reff@jnl{ZAp}}                
\def\nat{\reff@jnl{Nature}}             

\shorttitle{Bayesian methods of astronomical source extraction}
\shortauthors{Savage et al.}
\begin{document}
\title{Bayesian methods of astronomical source extraction}
\author{Richard S. Savage and Seb Oliver}
\affil{Astronomy Centre, University of Sussex, UK}
\email{r.s.savage@sussex.ac.uk}
\begin{abstract}
We present two new source extraction methods, based on Bayesian
model selection and using the Bayesian Information Criterion (BIC).  The
first is a source detection filter, able to simultaneously detect
point sources and estimate the image background.  The second is an
advanced photometry technique, which measures the flux, position (to
sub-pixel accuracy), local background and point spread function.  

We apply the source detection filter to simulated Herschel-SPIRE data
and show the filter's ability to both detect point sources and also
simultaneously estimate
the image background.  We use the photometry method to analyse a simple
simulated image containing a source of unknown flux, position and
point spread function; we not only accurately measure these
parameters, but also determine their uncertainties (using Markov-Chain
Monte Carlo sampling).  The method also characterises  the nature of the
source (distinguishing between a point source and extended source).

We demonstrate the effect of including additional prior knowledge.
Prior knowledge of the point spread function increase the precision of
the flux measurement, while prior knowledge of the background has only
a small impact.  In the presence of higher noise levels, we show that
prior positional knowledge (such as might arise from a strong
detection in another waveband) allows us to accurately measure the
source flux even when the source is too faint to be detected directly.

These methods are incorporated in SUSSEXtractor, the source extraction
pipeline for the forthcoming Akari FIS far-infrared all-sky survey.
They are also implemented in a stand-alone, beta-version public tool that can be
obtained at http://astronomy.sussex.ac.uk/$\sim$rss23/sourceMiner\_v0.1.2.0.tar.gz
\end{abstract}
\section{Introduction} \label{introduction}
Source extraction is close to ubiquitous in
modern observational astrophysics.  The ability to identify and
accurately quantify objects of interest in astronomical observations,
in particular with reliable automated methods, is becoming ever more
important with the advent of modern, large-area surveys.  It is
crucial that we are able to ask precise, statistical questions of the
data from these surveys.  Is there a source at a given
location in the sky?  Is it point-like or extended?  And what set of
parameters can define it?   Any science
derived from the study of astronomical objects proceeds directly from
accurate source extraction.  

In order to extract sources from astronomical data, we typically face
a number of challenges.  Firstly, there is instrumental noise.  It
is often possible to measure this instrumental/observational
characteristic and use this information to partially offset the
effects. More problematic are any so-called 'backgrounds' to the
observation.  These can be due to galactic emission, cosmological
backgrounds, faint source confusion, or even simply emission from
parts of the telescope itself.  These are often much harder to account
for and often constitute an in-depth study in themselves.  A prime
example is the extraction of sources from Cosmic Microwave Background
(CMB) data \citep[see e.g.][]{Vielva-01}.  We may also have to contend
with systematic effects such as glitches that can be caused by cosmic
ray hits on the detectors of space telescopes.

Because of these challenges and also because it is critical to exact
the utmost precision from our (often very expensive to gather) data,
we must strive to use all the available information when extracting
sources.  This means not only using all available data samples, but
also accurate noise estimates, measurements of the point spread
function and also inclusion of any other prior knowledge we may have.

Over the years, a number of methods have been created in order to use
various sets of information to obtain 'optimal' (subject to certain
sets of assumptions) source extraction methods.  There are many
techniques based on the concept of filtering data to enhance
relatively the signal due to objects of a certain set of
characteristics.  Examples of these include the matched, scale
adaptive and wavelet filters \citep[see e.g.][]{Vio-02, Barnard-04,
  Lopez-05, Barreiro-05}.  More recently, \citet[][]{makovoz-05} have
derived a filter of this type using the Bayesian formalism, thus
allowing for the explicit inclusion of prior knowledge.

Fitting of the point spread function to image data has also been used
as a way of accurately determining the position and flux of a (point)
source, \citep[see e.g.][]{Scott-02} The model-fitting methodology has
been given a much more general grounding in statistical theory by
\citet[][]{Hobson-03} who have detailed a very general (and powerful)
Bayesian framework for the extraction of sources.  Bayesian
methodology has also used in the Poisson noise regime \citep[see
  e.g.][]{Guglielmetti-2004}.
There are a number of publicly available source extraction packages,
which use a variety of the above methods (plus some other measures) in
order to accurately extract sources.  These include, for example,
DAOPHOT \citep[][]{DAOPHOT-87} and Source Extractor
\citep[][]{SExtractor-96}.

Perhaps the most flexible of these approaches is that of using
Bayesian statistics \citep[see e.g.][]{JaynesStats-book,
  MackayInfoTheory-book}, as it allows one to ask very precise
statistical
questions of the data. This framework is also highly general, allowing
the inclusion of all pertinent information.
In this paper, we explore the use of Bayesian statistics for source
extraction.  We present a pair of new methods based on this formalism,
one for simultaneous source detection and background
estimation/subtraction, and the other for an advanced form of source
photometry that also allows the determination of the nature
(point-like, extended etc) of the source.  

The contents of this paper are therefore as follows.
In \emph{Section \ref{section:methods}}, we present a general
  formalism for performing Bayesian source extraction.  We also detail
  two specific implementations.
In \emph{Section \ref{section:results}} we apply these methods
  to a simulated data sets, in order to demonstrate their abilities.
Finally, our conclusions are presented in \emph{Section
    \ref{section:conclusions}}.
\section[]{Methods} \label{section:methods}
In this section, we present a general formalism for performing
Bayesian source extraction.  We then apply this formalism to derive
two specific source extraction methods, with an eye to the analysis of
modern, large photometric astronomical surveys (although their
applicability is more general).  For this reason, both methods will
address 2D (i.e. photometric image) data, although we note that the
formalism extends to an arbitrary number of data dimensions.

Classic source extraction methodology divides the overall task into
two distinct stages, source \emph{detection} and source
\emph{photometry}.  
While the Bayesian paradigm allows for the possibility of a
single, combined approach, the nature of the data we are considering
dictates that we resist this.
Modern photometric surveys are often
large enough that such a combined approach is likely to be
computationally prohibitive.  The methods we present below retain the
two-stage approach, thereby proving computationally much quicker to
use.  

We note that in the following subsections, we will assume throughout
that the noise on each image pixel is Gaussian, of known variance, and
uncorrelated from pixel to pixel.  Additionally, when we are summing
over pixels, we will always choose a subset of the image pixels that
are local to the centre location we are considering.   A method for
determining optimally such subsets is given in \ref{subsection:dataSubset}.

The assumption of Gaussian noise warrants some discussion. 
In many real applications the noise distribution will naturally be 
close to Gaussian, e.g. when the dominant noise comes from well behaved
instrumental noise. In other cases a Gaussian 
 distribution might be inappropriate, e.g. in an
context where the data are strictly non-negative.  In some such cases a Poisson distribution might provide
a more natural description, when the photon statistics dominate. However,
if the photon numbers are sufficiently high then a Gaussian model 
is an adequate approximation to the Poisson distribution.  This condition
arises often in astronomy, e.g. when the sky background dominates.  In
 the case-study we are considering, observations with Herschel, the noise is dominated by the thermal background of the warm telescope primary and the Gaussian 
 approximation is reasonable. It would be possible to generalise the method 
 to include non-Gaussian noise distributions, including Poisson or log-Normal 
 distributions but that investigation is beyond the scope of this paper.

\subsection{General formalism}
The essence of Bayesian data analysis is to create a reasonable
parameterised model of the data.  These parameters can then be
constrained by the data themselves, along with any available prior
knowledge.

We begin with Bayes theorem.

\begin{equation}
P(\theta|D,H)=\frac{P(D|\theta,H)P(\theta,H)}{P(D|H)}\,,
\label{BayesTheorem}
\end{equation}

Where $P(\theta|D,H)$ is the posterior probability of the model
parameters ($\theta$), given the data $D$ and a hypothesis $H$.
$P(D|\theta,H)$ is the likelihood of the data (henceforth referred to
as $\cal{L}$, for simplicity) given a set of model parameters, $P(\theta,H)$
represents any prior knowledge we may have about the likely values of
the parameters, and $P(D|H)$ is the Bayesian Evidence.
Bayes theorem provides the framework for our work.  

We start with the likelihood.  If we are able to assess this, then
(after applying a prior), we will have the posterior probability
distribution, which is the result we require.  Following the normal
route for uncorrelated, Gaussian noise, we have the following:

\begin{equation}
\label{eqn:likelihood}
{\cal{L}} \propto \exp\left(-{\chi^2\over2}\right)
\end{equation}

\begin{equation}
\label{eqn:chiSquared}
\chi^2=\sum_{i=1}^{N_{pixels}}\left({{d_i-m(\theta)_i}\over\sigma_i}\right)^2
\end{equation}

and $d_i$ is the value of the $i^{th}$ data pixel of the subset of
image pixels under consideration, $m_i$ is the corresponding value
from a (parameterised) model of the signal and $\sigma_i$ is the
standard deviation of the (Gaussian) noise associated with that pixel.

Calculation of the likelihood function therefore depends on the
parameterised model of the signal that we are considering.  In this
case, the model will contain a source (point-like or extended).  It will also contain a
representation of the astronomical/instrumental background, as well as
possibly containing parameters describing instrumental characteristics
For example, if not well-defined by independent measurements, the
point spread function could be parameterised, and hence simultaneously
measured by the model-fitting procedure.

Once the likelihood has been constructed, any prior knowledge that we
have about the parameters can be included, in the form of the prior
probability (a density function spanning the same parameter space as
the likelihood).  This function might typically include information
such as prior knowledge of the positions of sources etc, although it
is perfectly acceptable to use an uninformative flat prior (i.e. equal
valued at all points in parameter space), if one has no relevant prior
knowledge (we note that this is the implicit assumption in maximum
likelihood methods).

As the Evidence is a constant, normalising term, we now have the
(unnormalised) posterior distribution.  We can map this distribution
by calculating posterior values over a hypercube of
parameter-space points or Markov-Chain Monte Carlo (MCMC) sampling.
The peak of this distribution is our most
likely solution, and (once normalised) the distribution as a whole
provides the statistical confidence regions.  

The posterior probability distributions of individual parameters 
can be obtained by marginalising over the other parameters \citep[see
  e.g.][]{SiviaBayesian-book}.  This can be done in a number of ways.
If MCMC sampling has been used to map the posterior probability
distribution, then simply making a histogram of the samples using the
values of a single parameter automatically gives the corresponding 1D
marginalised distribution (a well-known and highly useful feature of sampling from the
posterior).  If one were considering only a small number (three or
fewer, say) of parameters then it may be feasible to calculate
posterior values over a hypercube of parameter points and then
marginalise numerically (although this is a very brute-force
approach).  Or one can assume a functional form for the posterior and
perform the marginalisation analytically.  One common choice for the
functional form is that of a multivariate Gaussian, which is often a
reasonable approximation to the posterior and is analytically
tractable.  It also has the advantage that it can be completely
specified by a parameter covariance matrix evaluated at the maximum
a posterior point.

The method gives a complete analysis, given a particular choice of
model.  However, the question of selecting a good model still remains.
This can be addressed by the Evidence, which provides a relative
measure of the probability of different models being the best-fit,
given the data \citep[see e.g.][]{JaynesStats-book}.

Bayesian Evidence is typically time-consuming to calculate.  This
makes analytic approximations desirable, in terms of practicality.  In
particular, the Bayesian Information Criterion
\citep[BIC,][]{Schwarz-78} provides a easily calculated approximation to
the log(evidence).

\begin{equation} 
\label{eqn:BIC}
BIC = -2 \ln ({\cal{L}}_{max}) + \nu \ln (N_{data}) 
\end{equation}

Where ${\cal{L}}_{max}$ is the maximum likelihood value for a given
hypothesis, $\nu$ is the number of free parameters in the model and
$N_{data}$ is the number of (approximately equally weighted) data
used.  When comparing how likely different models are, lower BIC
values indicate higher probability of the model being the correct one.

Using model-selection criteria allows us to address the question of
which from a range of models is the best description of the data, and
to do so in a statistically rigorous way.  This becomes vital when
one's data contains millions of sources, some point-like, some
extended (and with different morphologies), and some not real at all,
but rather the product of contamination.

\subsection{Implementation: Bayesian source detection filter}
The first implementation that we present of the above formalism is a
Bayesian source detection filter.  
Source detection is necessary if one has observations of a region of
sky but has no explicit knowledge of the positions of sources in the
image (the case with many astronomical surveys).  Our task is
therefore to analyse the entire image, identifying the positions where
it is likely that there is a source present.

One consideration which is often critical for such source detection is
speed of analysis.  Modern photometric surveys, in particular, often
produce many large images, necessitating source detection methods that
are computationally quick to apply.  With this in mind, we derive an
analytic Bayesian solution to determine the relative probability (at
each pixel position in an image) of the data being best described by
an empty sky or a point source (with an unknown, uniform background in
each case).

The two models we therefore consider are the following.

\begin{description}
\item{\tt Empty sky, uniform background.}
This model consists solely of a flat, uniform background, described by
a single parameter (the level of the background).
\item{\tt Point source, uniform background.}
This model builds on the empty sky model, adding a single point
source, centred at the pixel currently being considered.  The point
source is modeled as a circularly-symmetric 2D Gaussian profile of
known FWHM.  This model has two parameters: the background level and the
integrated flux of the source.
\end{description}

We will compare these models using BIC.  This means (see Equation
\ref{eqn:BIC} that we only need to calculate the maximum posterior
value for each model.  By doing this at each (fixed) pixel position,
we can therefore calculate a map of the relative evidence for point
sources across the image.

Because we are considering (for each pixel) a fixed position, both
models are comprised of a linear sum of fixed components.  This means
that we can find analytic solutions in each case for the maximum
likelihood values.
Using the condition that the partial derivatives of the likelihood
must be zero at the maximum likelihood solution, we can solve to find
the following maximum likelihood solutions for each model.

For the point source model, we have the following description of the model.
\begin{equation} 
m_i = F {\cal{P}}_i + B
\end{equation}

For the empty sky model, we have the following simple description of the model.
\begin{equation} 
m_i = B
\end{equation}

Where $m_i$ is the $i$th model pixel, $F$ is the source flux, ${\cal{P}}_i$ is the (Gaussian)
point spread function (normalised such that it integrates to unity)
and $B$ is the uniform background.

>From this, we find the following analytic maximum likelihood solutions
for $F$ and $B$.

\begin{equation}
F_{source} = {{\gamma \beta - \delta \epsilon} \over {\alpha \beta - \epsilon^2}}
\end{equation}

\begin{equation} 
B_{source} = {{\alpha \delta - \gamma \epsilon} \over {\alpha \beta - \epsilon^2}}
\end{equation}

\begin{equation} 
B_{empty} = {\delta \over \beta}
\end{equation}

Where the calculated values used in the above equations are given by
the following (with all sums being performed over the image pixels in
a local region. See subsection \ref{subsection:dataSubset} for a
discussion of how to choose this region).

\begin{equation} 
\alpha = \sum_{i=1}^{N_{pixels}}\left({{{\cal{P}}_i^2}\over\sigma_i^2}\right)
\end{equation}
\begin{equation} 
\beta =  \sum_{i=1}^{N_{pixels}}\left({{1}\over\sigma_i^2}\right);
\end{equation}
\begin{equation} 
\gamma =  \sum_{i=1}^{N_{pixels}}\left({{d_i{\cal{P}}_i}\over\sigma_i^2}\right)\\
\end{equation}
\begin{equation} 
\delta =  \sum_{i=1}^{N_{pixels}}\left({{d_i}\over\sigma_i^2}\right);
\end{equation}
\begin{equation} 
\epsilon =  \sum_{i=1}^{N_{pixels}}\left({{{\cal{P}}_i}\over\sigma_i^2}\right)
\end{equation}

By then feeding the best-fit model back into Equations
\ref{eqn:likelihood} and \ref{eqn:chiSquared}, we obtain the maximum likelihoods.  We can
therefore calculate the relative BIC at each pixel position.
(we note that we implicitly use flat, uninformative priors in the
preceding steps)
The resulting map is an estimate of the (log of the) relative
probability of there being a point source, rather than empty sky, at
any given pixel position.  

The local extrema of this map therefore give us the locations where
one model is (locally) most favoured over the other.  Constructing the
map so that (by convention) high values correspond to the point source
model being more likely, we can identify the most likely source
positions in the input image by identifying the local maxima in the
map, subject to some minimum threshold value.

This method is closely modeled in some respects on the traditional filtering methods
such as matched, scale-adaptive and wavelet filters.  It does,
however, have several key advantages.
   
\begin{enumerate}
\item{\tt Simultaneous background estimation, subtraction.}  
  In real astronomical data, background subtraction is a highly non-trivial
  task.  In particular, more traditional methods such as median
  filtering are biased by the presence of sources.  By performing the
  subtraction simultaneously, we largely avoid this problem.
\item{\tt Proper accounting for flagged data and locally-varying noise.}
  Real astronomical images will typically have gaps due to flagging
  and uneven scan strategies, as well as point-to-point variations in
  noise levels.  This approach allows us to properly account
  for these effects by including an individual statistical weight
  (i.e. $1\over\sigma^2$) for each image pixel.  Similarly, setting
  a given weight to zero effectively flags out the corresponding datum
  (see Equation \ref{eqn:chiSquared}).  This is mathematically
  well-defined; the principle challenge in such cases is in fact to
  estimate accurately the statistical weight (via the standard
  deviation) for each image pixel, which will depend on exactly how
  the image was created (for example, if it is the sum of many
  repeated observations, the multiple samples contributing to each
  image pixel can be used to estimate the standard deviation).
\item{\tt Extensible method.}
   As it is based on a very flexible and general formalism, this
  source detection filter can be straightforwardly modified to accommodate more
  complex and realistic data models.  For example, many data are
  subject to 'glitches' (caused by cosmic ray hits
  on detectors).  By including in this method a third model of a
  single very high pixel value, it would be possible to distinguish
  between a source and a glitch-spike.

\end{enumerate}

\subsection{Implementation: Bayesian source photometry}
\clearpage
\begin{figure*}
\begin{minipage}{150mm}
\includegraphics[angle=-90, width=4.5cm]{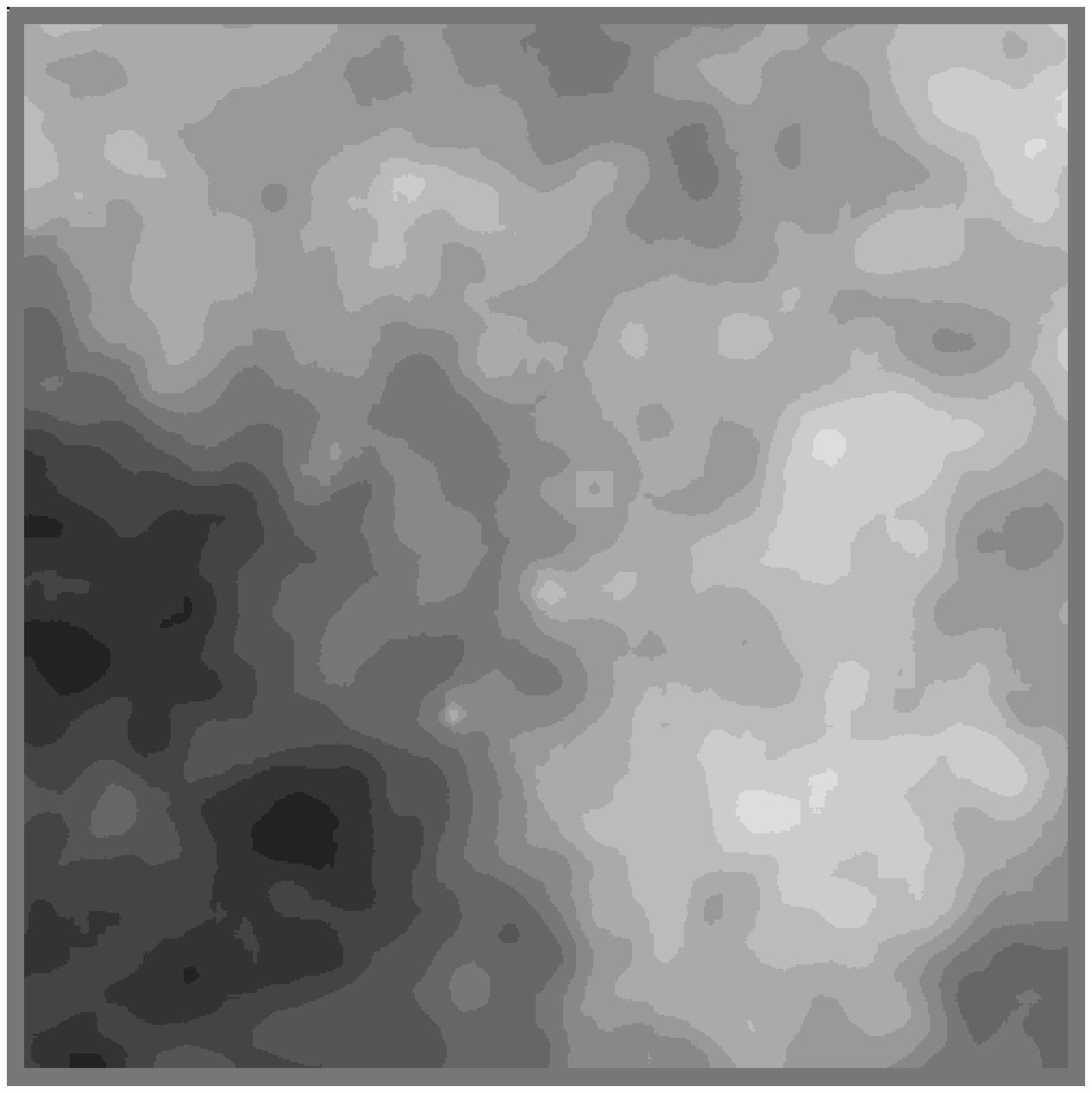}
\includegraphics[angle=-90, width=4.5cm]{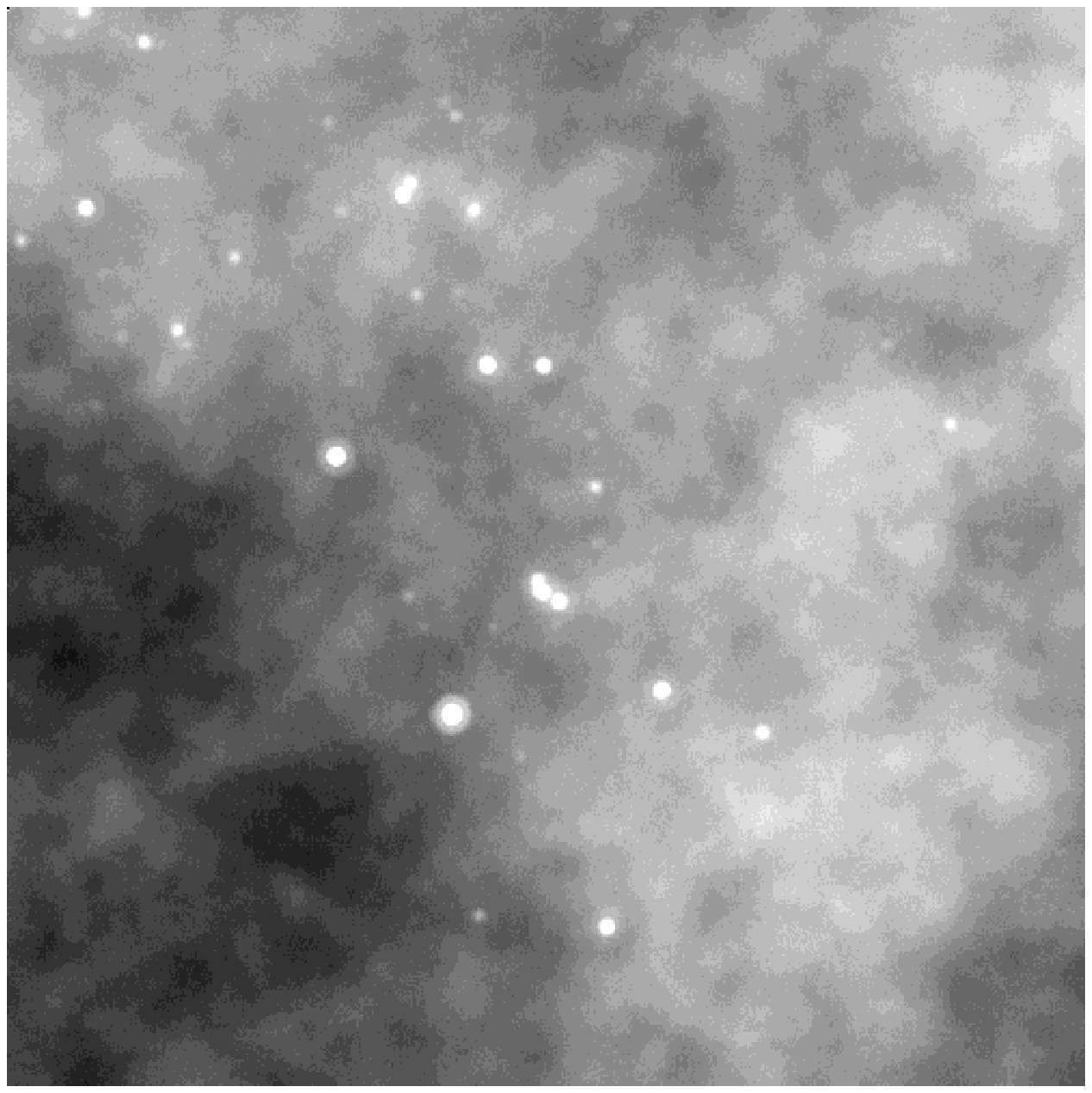}
\includegraphics[angle=-90, width=4.5cm]{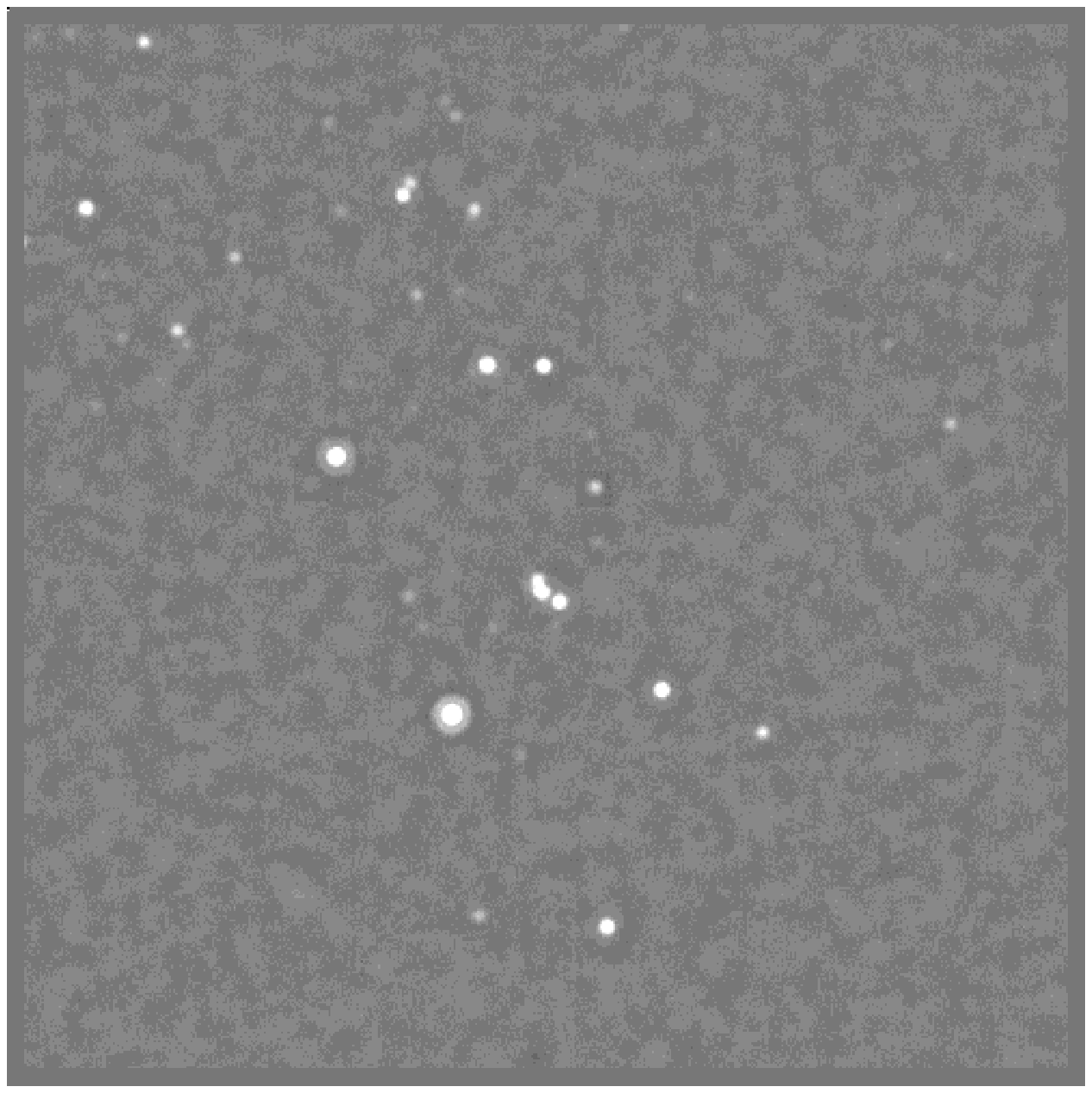}
\caption{The input flux image
  (left), a map of the background (as estimated by the source
  detection filter) (middle) and a map of the residuals between input
  flux image and background map (right) (with matched grey-scales).  As can be seen, even the
  presence of very bright sources does not appreciably bias the
  background estimation of our algorithm.  We also note that by
  setting a threshold of 550 (unusually high, due to small-scale
  fluctuations in the diffuse background, which act as a source of
  correlated noise) we are able to successfully detect 26 of the 33
  sources (with 3 false detections). Simulated flux image courtesy of
  Philippe Andre, Bruce Sibthorpe and Tim Waskett.}
\label{fig:sourceDetection}
\end{minipage}
\end{figure*}
\clearpage
Once a source has been detected, we wish to more completely measure
and characterise it.  Considering only regions of the sky in which
there is likely to be a source means that we can afford to devote
substantially more computational effort to each candidate position.
This is the principal advantage of performing source extraction in two
distinct stages.

In this method, we again adopt the approach of fitting multiple models
to the local data (again using the fact that we are interested in
compact sources to minimise the data we must consider).  However in
this case we will use a more in-depth approach, allowing more
parameter to vary and mapping out the posterior probability
distribution in each case.  The result will be more precise results
(in particular, sub-pixel positional accuracy) and the
determination of the errors on each parameter (without assumptions as
to the form of the error distributions).

We proceed again by defining a number of models which we will fit to
the data.

\begin{description}
\item{\tt Empty sky, uniform background.}
This model consists solely of a flat, uniform background, described by
a single parameter (the level of the background).
\item{\tt Point source, uniform background.}
This model builds on the empty sky model, adding a single point
source at a given (parameterised) X, Y position.  The point
source is modeled as a circularly-symmetric 2D Gaussian profile of
known FWHM.  This model has four parameters: the background level, X
and Y position and the integrated flux of the source.
\item{\tt Extended source, uniform background.}
This model is the logical extension of the point source model and is
identical, with the exception that the FWHM is now allowed to vary as
a model parameter (giving five in total).  This allows us to account
either for circularly symmetric extended sources, or alternatively to
measure the FWHM of  the point spread function, if this is not known.
\end{description}

We emphasise that there are many other models that can be usefully
applied.  Examples would be non-circular extended sources, models
where the noise is unknown or models where there are two or more
adjacent (blended) sources.

For simplicity, we will again reply on the BIC for model comparison,
although a full Bayesian Evidence calculation could be used (computing
resources permitting).

As before, the likelihood functions are thus defined for any given set
of parameter values of the relevant model.  Multiplying by the prior
distribution for each model, we have the posterior for each model,
which is mapped using MCMC sampling (except for the empty sky model,
for which we only require the analytic best-fit solution, unless a
prior is imposed).

The MCMC sampling returns the best-fit value for each model.  We use
this to calculate BIC values and hence determine which model is mostly
likely to be the best representation of the data.  This characterises
the nature of the source in question.

Returning to the MCMC samples for the most likely model, we have also
mapped the posterior probability distribution for that model.  From
this we can straightforwardly determine the confidence intervals and
best-fit values for all fitted parameters.

The power of this method lies in its ability to ask precise,
statistical questions as to the nature of a source and to recover the
theoretically optimal amount of pertinent information, given the
data.  The flexibility of the Bayesian framework means that we are
able to adapt this method, depending on the type of sources (and data)
that we are expecting.  It is entirely realistic to deploy a whole
battery of models, fitting each one in turn and determining which is
the most likely representation.

\subsection{Determining optimal data subset size} \label{subsection:dataSubset}
Because we are concerned with the extraction of compact sources, the
above analyses need only consider a small subset of \emph{local} data,
for each source position.  This region should be large enough that we
get good constraints on the source flux and local background, but
small enough that our assumption of a flat background does not break
down.  Our definition of the size and shape of this region will
therefore have a direct impact on our source extraction.

We choose to determine an optimal region size in terms of a minimised
BIC value (and hence maximised Bayesian Evidence).  This will give us
a data model that best describes our data (for the types of model we
are considering here).  If we define the region as circular, then we
reduce this problem to an optimisation (in BIC) with respect to the
radius of the region.  

One complication is that for BIC comparisons to be valid, we require the
same data set to be considered in each case.  This would plainly not
be true if we simply used the data inside the circular
\emph{region-of-interest}.  To
avoid this problem, we also define a larger super-region (also circular), 
and label the super-region image pixels that lie outside the
region-of-interest as \emph{external pixels}.  We then redefine our
model as fitting the source/background to the
\emph{region-of-interest}, plus allowing additional free parameters
for each of the \emph{external pixels}, so that they are fitted
exactly and do not contribute to the chi-squared of the model fit.
Therefore, the \emph{external pixels} will contribute to the BIC
solely as extra parameters, the number of which will vary depending on
the radius of the \emph{region-of-interest}.

These nuisance parameters are not trivial to deal with and we
highlight that the above procedure makes the simplifying assumption
that the nuisance parameters can be
fitted to the data with no uncertainty, so that marginalisation over
them is not necessary.  In practice this is not true and would alter
the BIC calculation (via the maximum likelihood value).  

This could be accounted for in the photometry method because MCMC
methods can straightforwardly include large numbers of nuisance
parameters, which can be marginalised over without extra effort.
Doing so will incur the need for longer sampling chains to be
generated, to ensure adequate convergence.

One peculiarity of this procedure is that the BIC has a weak
dependence (going as the logarithm) on the radius of the super-region
(and hence the maximum possible region radius).
While this is clearly undesirable, the effect will be small for reasonable
radius ranges.

The minimum sensible region radius will typically be dictated by the
FWHM of the point spread function, with perhaps a radius equal to the
FWHM being a reasonable starting point.  The maximum region radius is
less well-defined, but a value of four or five times the FWHM would
seem intuitively reasonable, and our experiences in this paper suggest
that this is not unreasonable.

For the case of fitting a single source (i.e. when we are applying source
photometry), this process is unambiguous. In the case of the source detection filter, where we may
have many detected sources (and that number may change as we optimise
with radius), we need to choose what metric we will optimise.  In
general, this choice will depend on the exact nature of both the data
and the science in question..  

One simple approach (and the one that we have adopted here) is to use
an intermediate radius case (say, twice the FWHM of the point spread
function) as a starting point, identifying the sources detected
in this case.  We then use as our metric the sum of BIC values for the
fits to these sources.

This procedure gives us a way of selecting an optimised definition for
a local region of the data.  This both optimises the performance of
the source extraction algorithm, but also means that the user need not
waste any time optimising by trial and error.

\subsection{Prior knowledge}
A strength of the Bayesian formalism is its explicit inclusion of
prior knowledge.  In the case of source extraction, one typically
assumes that the noise characteristics and the point spread function
are known (although this need not be the case).  One could also assume
prior knowledge about any of the fitted parameters; for example,
source position may have already been determined in another observing
band.  

It is also possible to select priors on the basis of more general
knowledge.  For example, if one is attempting to detect a population of galaxies,
it may be reasonable to assume a power law distribution for the source
flux (e.g. from a model of the galaxy population).  Even in
the absence of such knowledge, one could still choose the
\emph{Jeffreys prior} (a power law with index of $-1$), which is the
indifference prior for a positive-only scaling parameter.

The source flux is also of particular note because it will typically
(although not always; e.g. Sunyaev-Zel'Dovich effect for galaxy
clusters in CMB observations) be subject to the constraint of being
non-negative.  In this case, it is important to properly apply this as
a prior constraint. 

Because the above source photometry method uses MCMC
sampling, it is straightforward to quantify the assumptions on the prior.  In section
\ref{subsection:sourcePhotometry}, we show examples of this.

The source detection filter relies on analytic solutions in order to
give plausible speed of analysis.  This makes the application of
non-top-hat priors more difficult, if convenient analytic solutions
are to be possible.  The potential size of this topic takes it beyond
the scope of this paper, but we note that the exploration of different
priors represents a largely untapped area where source detection
methods could be improved.

\clearpage
\begin{figure*}
\begin{minipage}{150mm}
\includegraphics[angle=0, width=8cm]{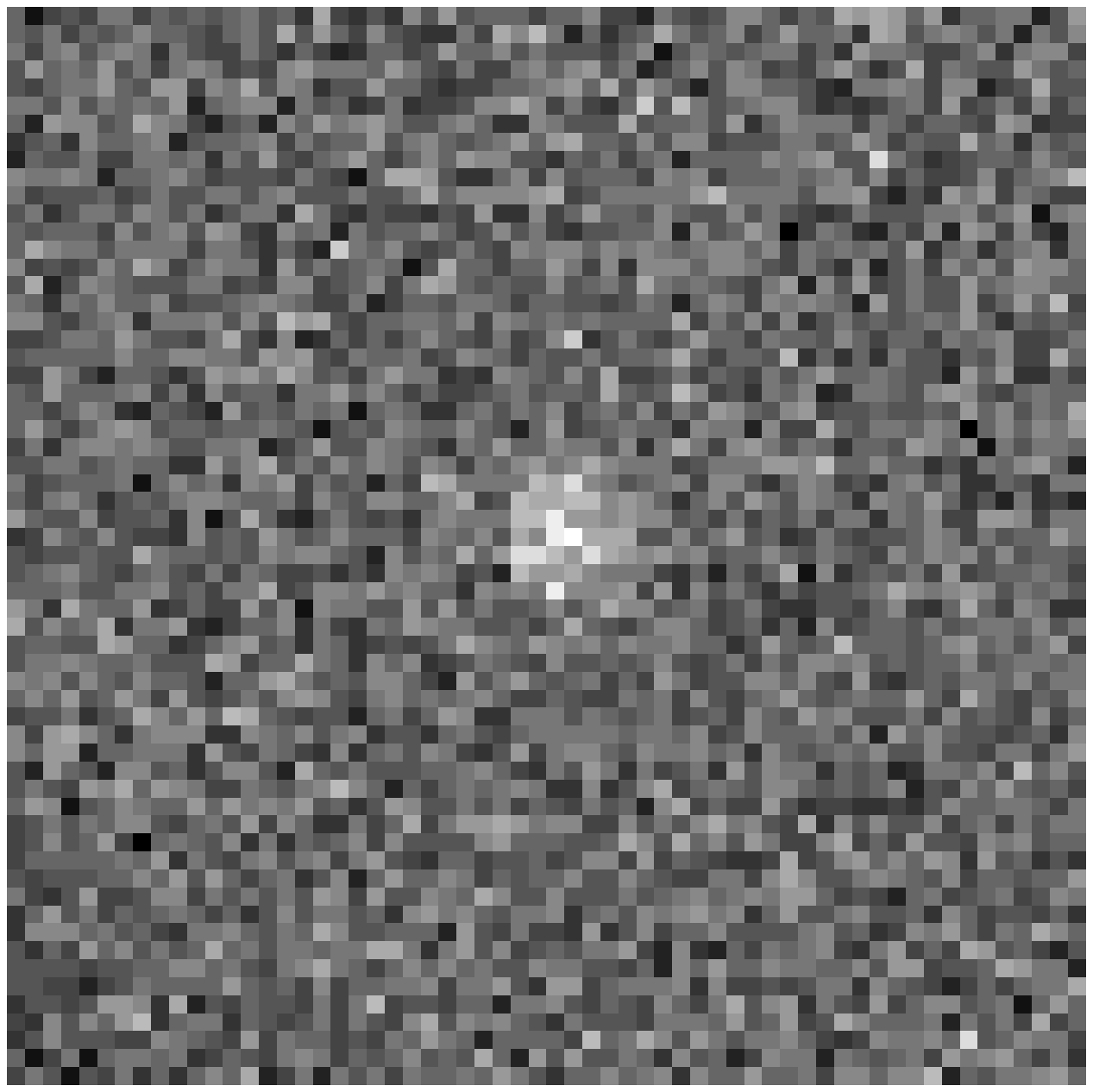}
\includegraphics[angle=0, width=8cm]{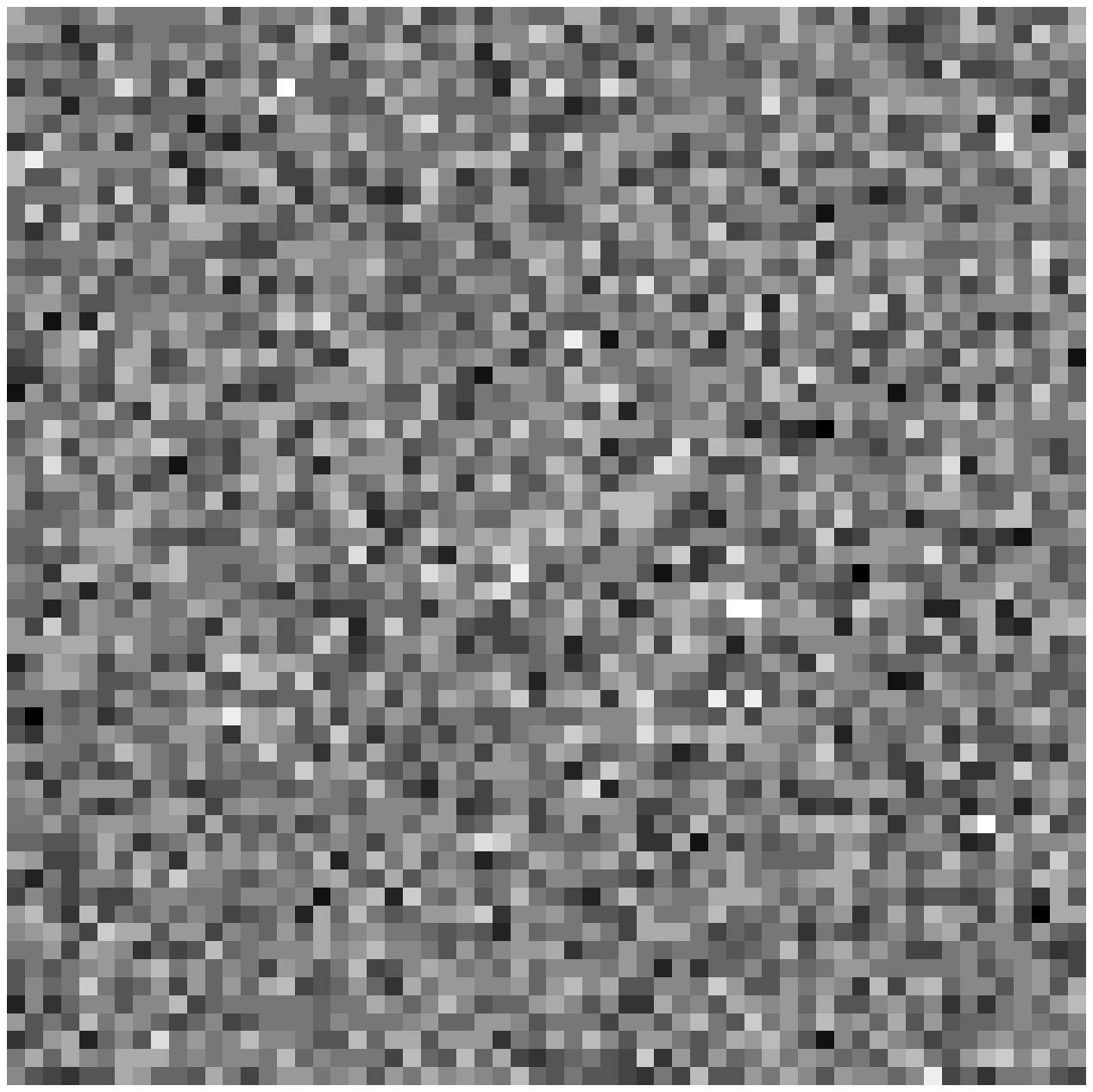}
\caption{The simulated images on which photometry was performed.  Both
images contain the same underlying signal, consisting of a uniform
background (of level 0.5 units), plus a Gaussian point source with
position (relative to the image centre) of (0.3, 0.4) pixels, a FWHM
of 5.2 pixels and in integrated flux of 10 units (corresponding to a
peak height of 0.231 units).  The left-hand image has Gaussian noise
with rms of 0.075, the right-hand has rms noise of 0.3 (i.e. higher
than the peak of the source).}
\label{fig:simulatedData}
\end{minipage}
\end{figure*}
\clearpage

\section[]{Results} \label{section:results}
In this section, we present example results from the two methods
detailed in the previous section.  We highlight the speed of analysis
of these methods.  Running on a desktop machine (using two 2.4~GHz AMD
Opteron 250 CPUs) and implemented in IDL, the source detection filter
processed $9 \times
10^4$ pixels per second (a $784 \times 912$ pixel image in eight
seconds), and the photometry method was able to analyse one source
every nine seconds (producing $10^5$ MCMC samples per model, per
source).  At this rate, for example, the whole Akari all-sky survey could be source
detection filtered in four days and four hours (assuming $40,000$
square degrees of coverage, with $8 \times 8 arcsecond$ image pixels
and four observing bands, using a single desktop machine.   

\subsection{Bayesian source detection filter}
Figure \ref{fig:sourceDetection} shows images from the analysis of a
simulated Herschel-SPIRE \citep[see e.g.][]{HerschelOverview-2004} observation of a number of point sources,
along with a diffuse galactic foreground (data courtesy of Philippe
Andre, Bruce Sibthorpe and Tim Waskett).  Shown are the input flux image, a map of the
background (as estimated by the source detection filter), and a map of the
residuals between input flux image and background map.

The background map is created using the maximum a posterior estimate
of the model background at each pixel position.  In each case, the
model used is that which is most likely, on the basis of BIC score.

The residuals map is created by subtracting the background map from
the original input flux image.  The residuals will therefore contain
the point sources, plus any imperfections in the background
estimation.

\subsection{Bayesian source photometry} \label{subsection:sourcePhotometry}
\clearpage
\begin{figure*}
\begin{minipage}{150mm}
\includegraphics[angle=0, width=8cm]{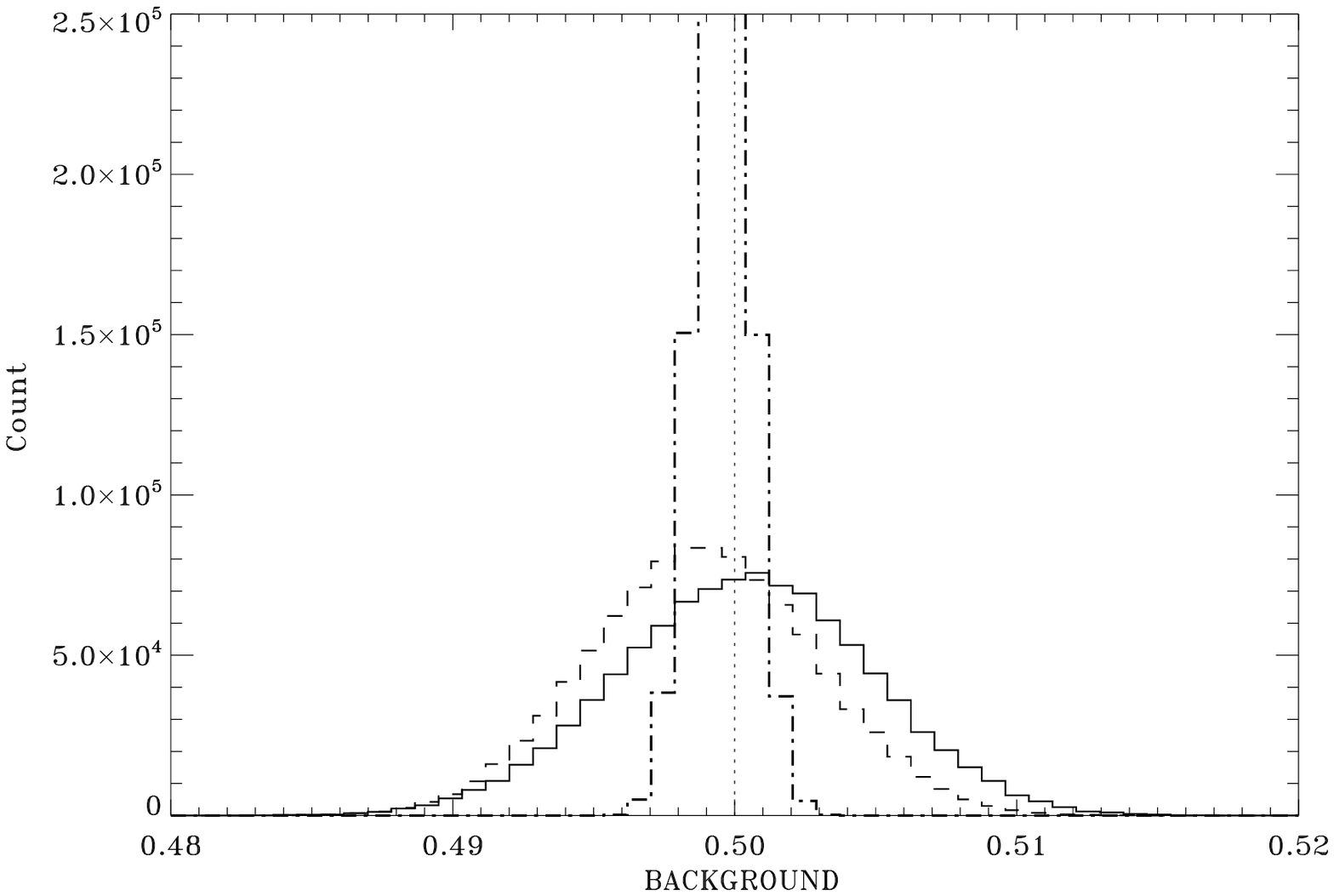}
\includegraphics[angle=0, width=8cm]{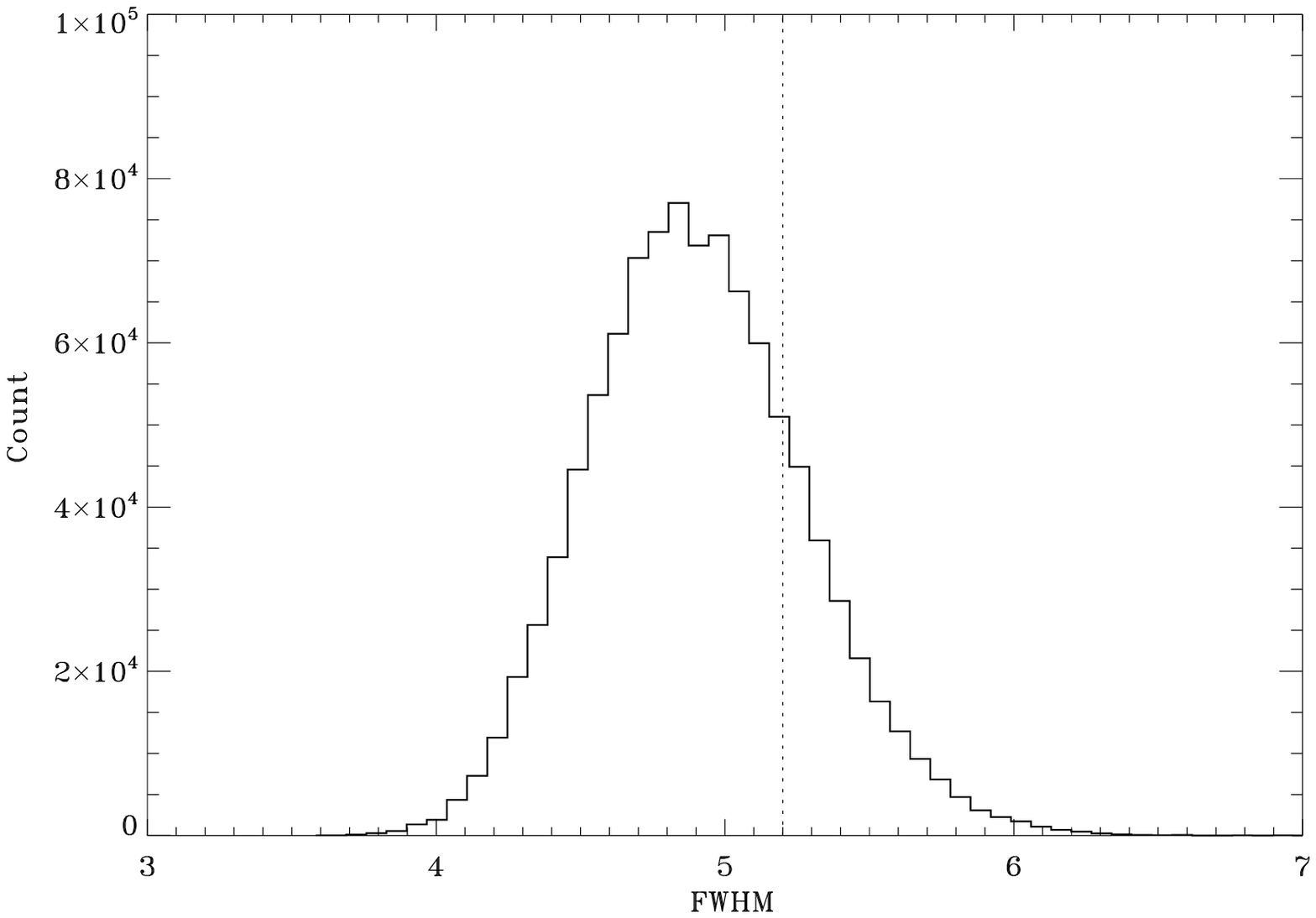}
\includegraphics[angle=0, width=8cm]{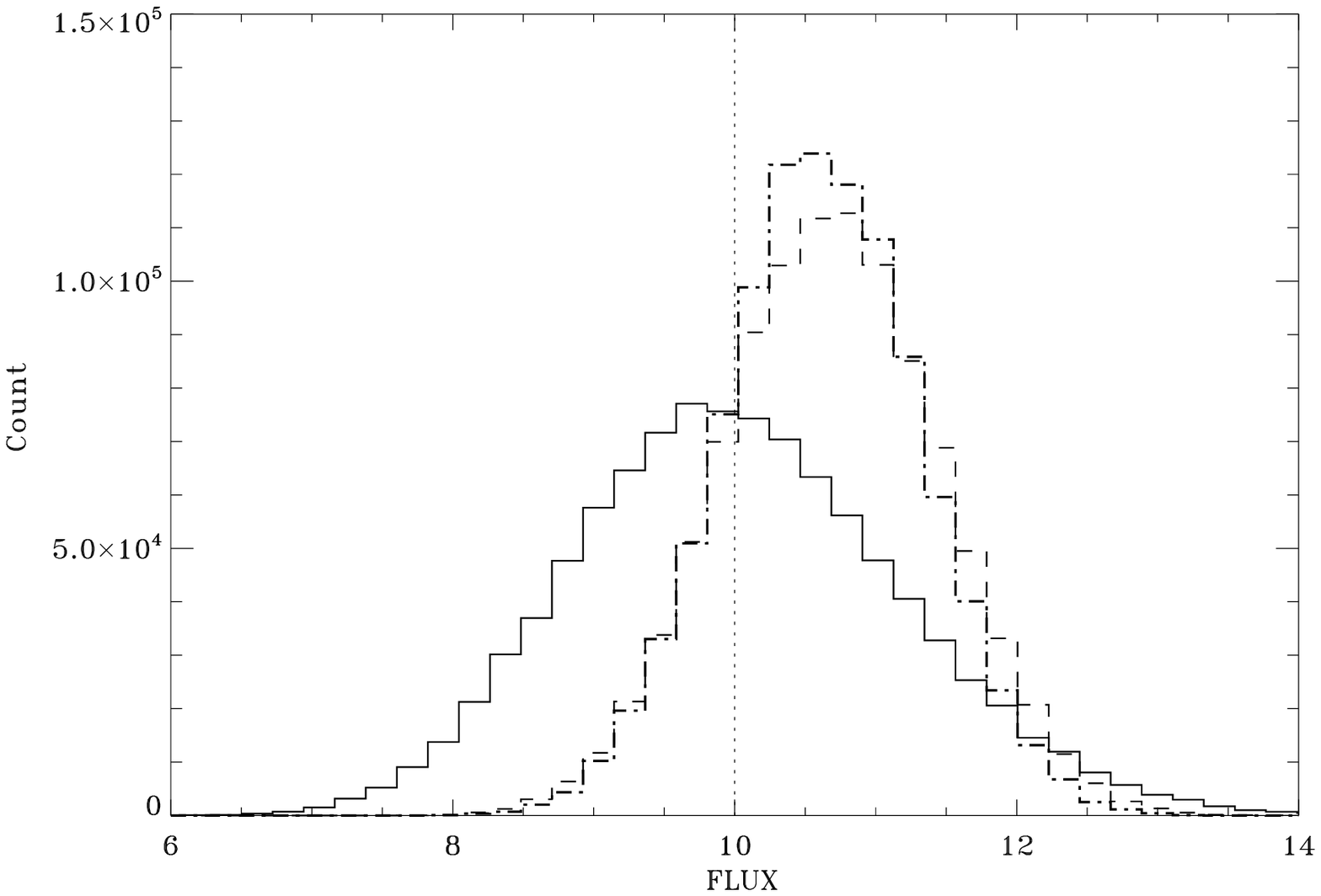}
\includegraphics[angle=0, width=8cm]{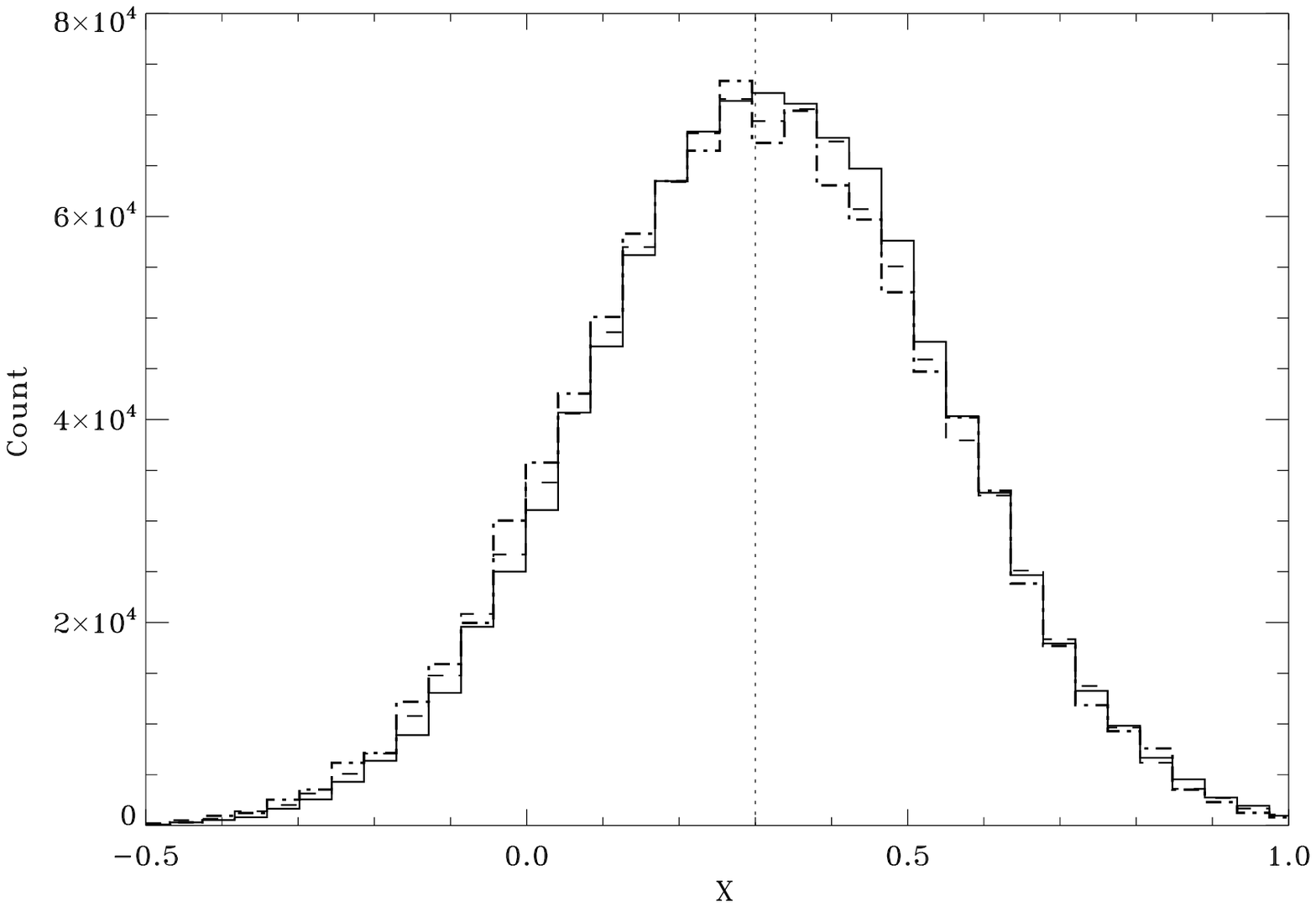}
\includegraphics[angle=0, width=8cm]{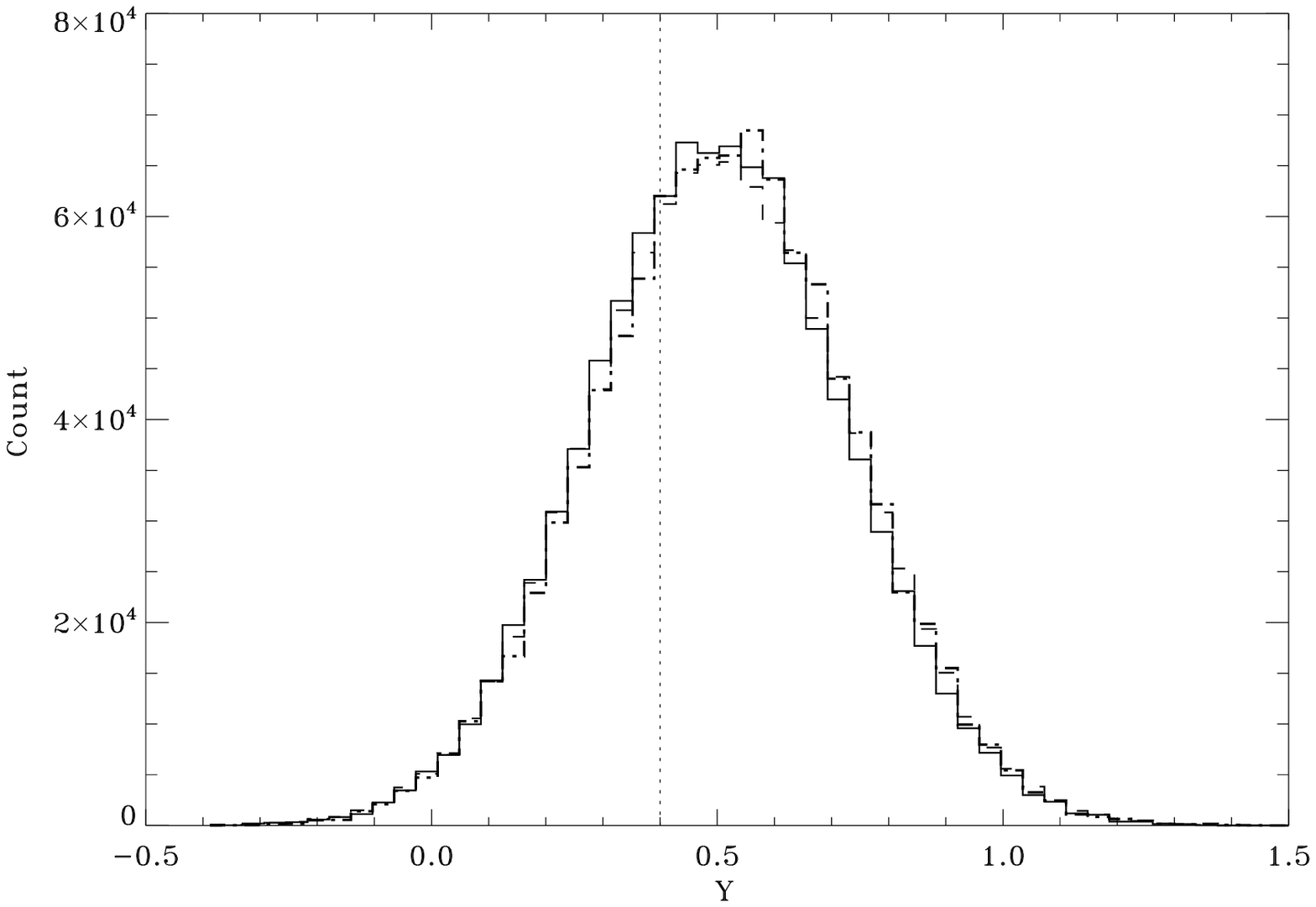}
\caption{1D marginalised posterior probability distributions for the
  five parameters of the extended source photometry model (solid
  line).  Also shown are the case where the FWHM prior is known perfectly
  (dashed line) and where both the FWHM is known and there is a
  Gaussian prior on the background (dot-dash line).  The input values
  are marked by vertical, dotted lines.}
\label{fig:compactSourcePosterior}
\end{minipage}
\end{figure*}
\begin{figure*}
\begin{minipage}{150mm}
\includegraphics[angle=0, width=8cm]{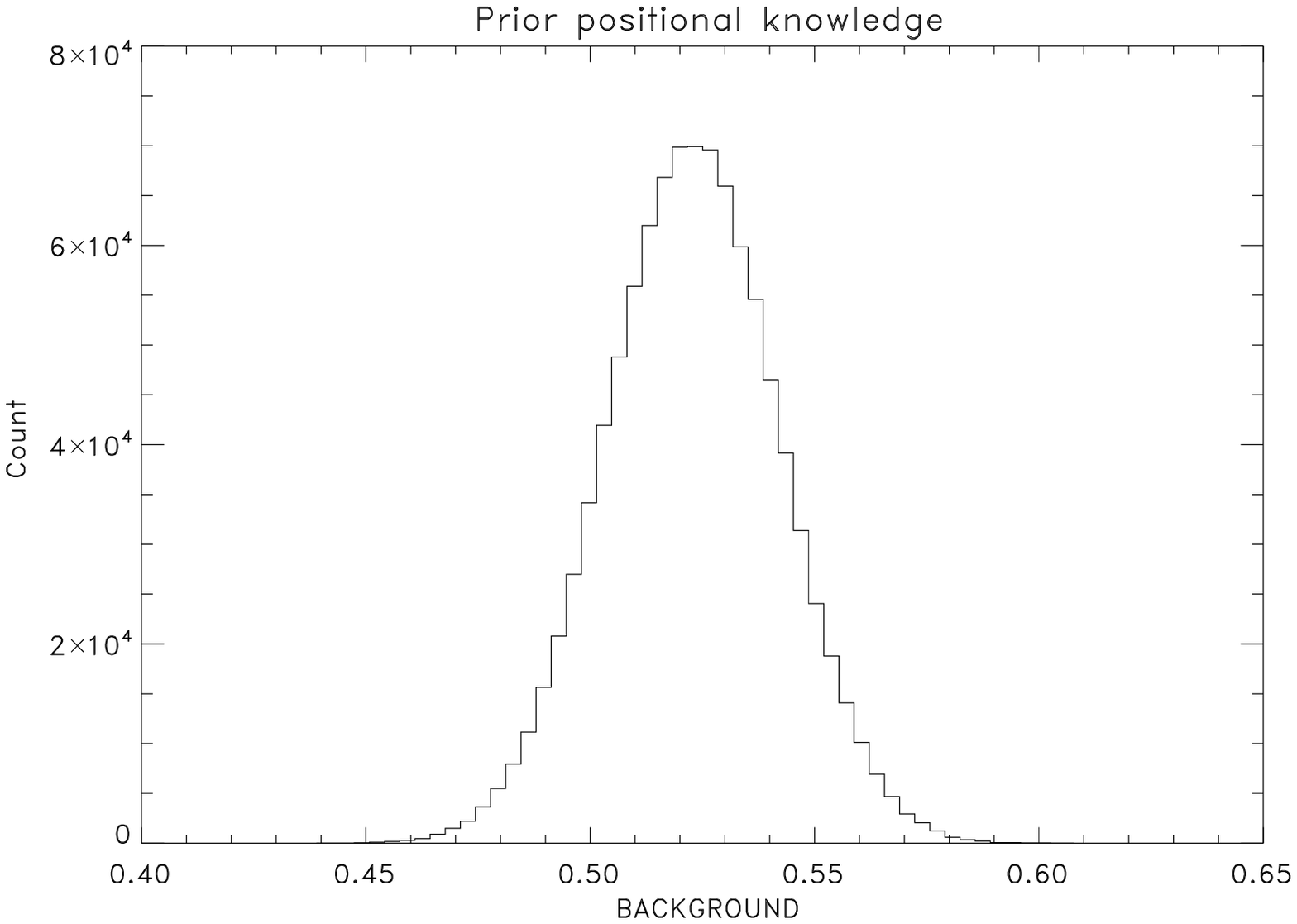}
\includegraphics[angle=0, width=8cm]{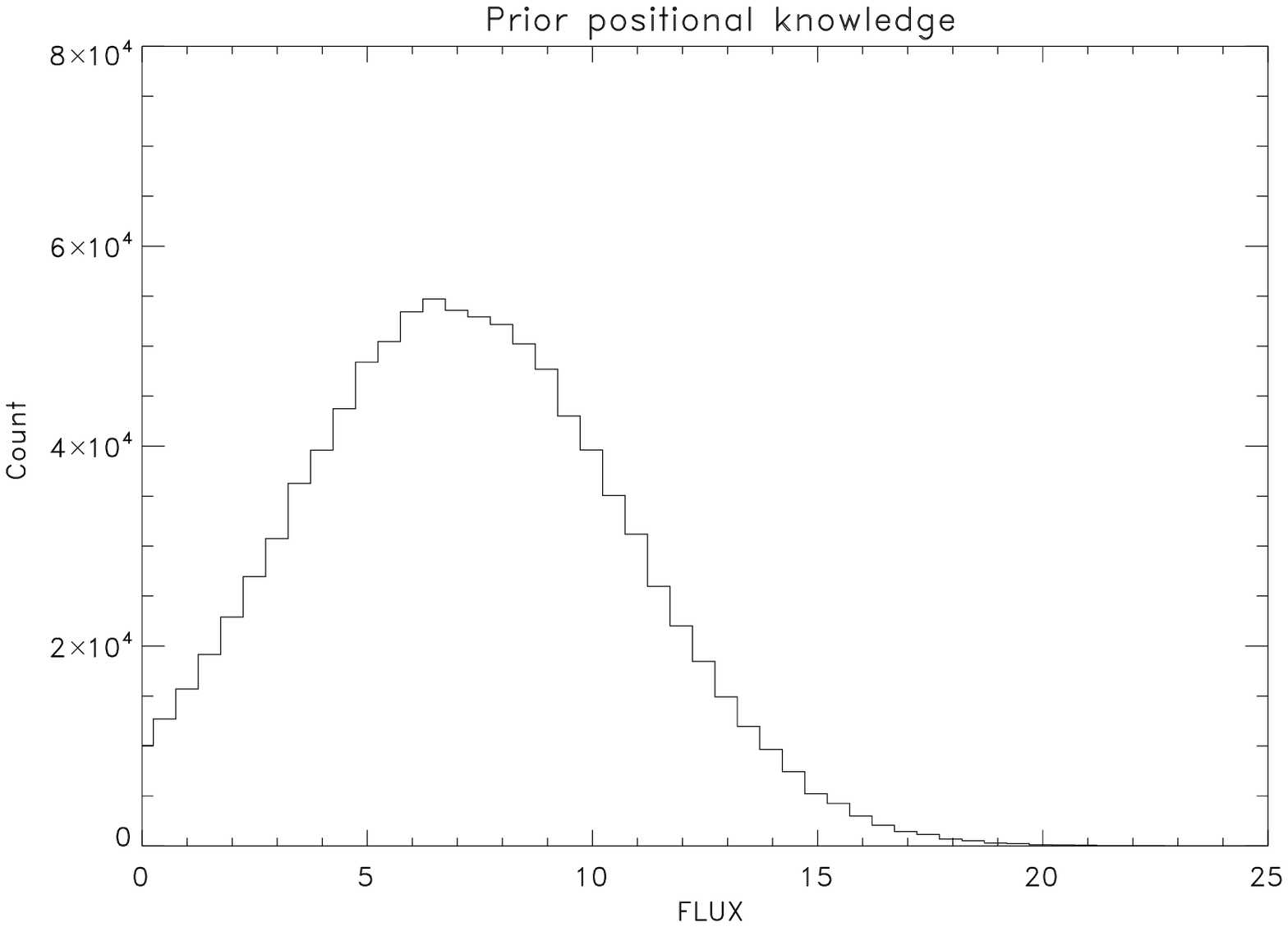}
\caption{1D marginalised posterior probability
  distributions for two parameters of the point source
  photometry model. Prior positional knowledge has been included, in
  the form of a Gaussian prior on both X and Y (FWHM of 0.1 pixels)
  In this case, the rms noise of the observation has been increased
  four-fold, so that in the absence of the prior, the BIC value would
  favour an empty sky.  This shows the case where a source has been
  detected to high precision in another band, but is very faint in
  this band.  The Bayesian formalism allows us to fully and properly
  account for this.}
\label{fig:positionPriorPosterior}
\end{minipage}
\end{figure*}
\begin{figure*}
\includegraphics[angle=0,width=16cm]{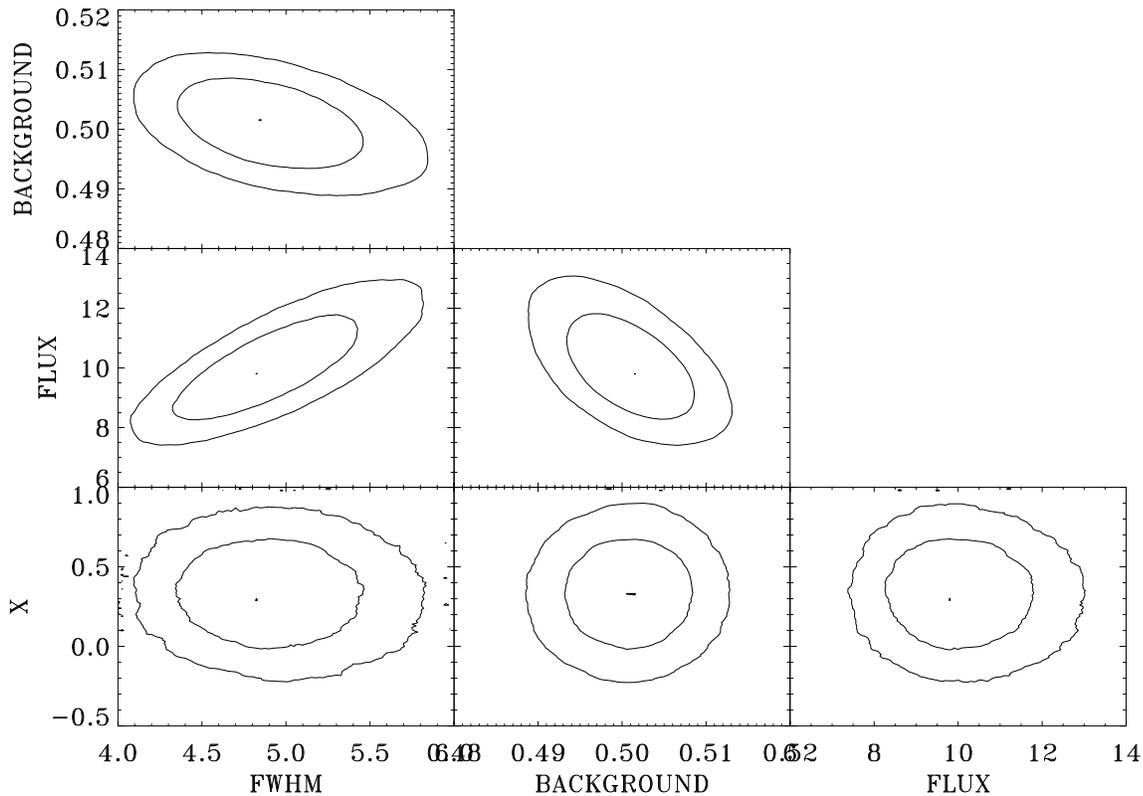}
\caption{Contour plots of the 2D marginalised posterior probability
  distributions for the parameters of the 'compact' source
  photometry model (shown are the 68\% and 95\% confidence regions,
  plus the maximum a posterior point).  The contours are found using smoothed 2D histograms of
  the MCMC samples.  This gives estimates of the marginalised 2D
  posterior probability distributions for these parameter
  combinations.  Note that we exclude the 'Y' parameter, as its
  behavior simply mimics that of 'X'.  The FLUX and FWHM of the point
  spread function are correlated.  The FLUX and BACKGROUND are
  negatively correlated.  The FWHM and BACKGROUND also show a slight
  negative correlation.  As expected, the X position is uncorrelated
  with these three other parameters.}
\label{fig:2Dposterior}
\end{figure*}
\clearpage
The analyses in this sub-section are carried on on a simple, simulated
test image (shown in Figure \ref{fig:simulatedData}).  
The image contains a single point source on a uniform background, with
uncorrelated Gaussian random noise added to each pixel.  While this is
a benign data set, it is instructive to consider such an idealised
case in order to better understand features of the algorithm.

Figure \ref{fig:compactSourcePosterior} shows the 1D marginalised
posterior probability distributions for a variety of cases.  The solid
lines show a five parameter 'compact'
source model fitted to the data.  The five parameters are a flat
background, the FWHM of a Gaussian point spread function, the flux of
the source, and its X and Y co-ordinates within the image.
The dashed lines show the 1D marginalised
posteriors for the case where the FWHM of the point spread function is
known(for example, it has been measured independently of this
'observation').  The dotted lines show the 1D marginalised
posteriors for the case where the FWHM of the point spread function is
know \emph{and} we have prior knowledge of the level of the
background.
Figure \ref{fig:positionPriorPosterior} shows the 1D marginalised
posteriors arising from analysing the same source with four times
the rms Gaussian noise.  The FWHM is taken as known, as is prior
knowledge of the the source position.  This simulates the case
where a source has been strongly detected at another band and we now
wish to find an estimate of that source's flux in this band.
Figure \ref{fig:2Dposterior} shows examples of 2D marginalised
posteriors for the five parameter 'compact' source model.  These
illustrate the different correlations that exist between the fitted
parameters.

\section{Conclusions} \label{section:conclusions}
In this paper, we have described a Bayesian formalism for the
extraction of sources from astronomical data and have used it to
derive two new source extraction methods.  We then demonstrated the
methods on simulated data.

The source detection filter is a deliberately uncomplicated
implementation of this formalism; it is designed to analyse images
quickly, something that is often crucial given the size of many modern
astronomical surveys.  Estimation of the image background is an
often-overlooked (and highly non-trivial) aspect of source extraction
and the simultaneous estimation performed by our filter makes unbiased background
subtraction much more tractable.  An additional point not to be
under-estimated is that by combining background subtraction and source
detection, we have created a method that has essentially only one
user-defined parameter (threshold), substantially simplifying its use.  

We applied this filter to a deliberately challenging simulated image.
The presence of a strong diffuse astronomical background introduces
fluctuations on similar angular scales to the point spread function,
presenting a particular challenge for source extraction. In spite of
this, we are still able to detect the majority of sources, with only a
few spurious detections.  If computationally fast ways can be found to
better model this background (work beyond the scope of this paper),
even more impressive results may be possible in the future.

Once a candidate source position has been identified, we wish to
characterise the source as precisely as possible. The advanced
photometry method allows us to do just that.  It can
determine the flux, position (to sub-pixel accuracy), local background
and (if required) point spread function FWHM, along with the
uncertainties on those estimates.  Furthermore, it allows the meaningful
comparison of different models, allowing us to determine (in an
automated way) whether any given source is point-like, extended or
even just a patch of empty sky.  We can also include any additional prior
knowledge we may have about the source. For example, if the FWHM is
known then the precision of our flux estimate is improved.  With
prior positional knowledge (from a strong detection in another band),
we can obtain a flux estimate even when there is insufficient evidence
from the data alone to identify a source.

This formalism allows us to ask precise, statistical questions of our
data.  We are able to include all pertinent information, giving us the
best possible measurement and characterisation of the sources.  We
can also determine a number of figures-of-merit, such as Bayesian
Evidence, BIC and reduced chi-squared, all of which give measures
of the quality of the extraction.  Parameter
space searching techniques such as MCMC sampling allow us to recover
the statistical uncertainties on our measurements while making
minimal assumptions.  And model selection techniques allow us to ask
which of a range of models best characterise any given source.

In conclusion, in this paper we present the following.
\begin{enumerate}
\item{\tt A Bayesian formalism for the detection and extraction of
  compact sources from astronomical data}
\item{\tt The derivation of an analytic source detection filter that
  simultaneously detects point sources and estimates the image background}
\item{\tt The detailing of an advanced photometry method, which
  determines source parameters such as flux and position (to sub-pixel
  accuracy), as well as their uncertainties. It also allows us
  to determine the nature of the source (point-like, extended) and to
  include any prior knowledge we may have, thus enhancing the precision
  of our results}
\item{\tt A method for optimising the local region from which data
  should be used to make the source fits} 
\end{enumerate}

Bayesian source extraction is a highly powerful and (perhaps
just as importantly) immensely flexible methodology.  The ability to adapt our
methods to the peculiarities of the data we are considering is a key
degree of freedom when dealing with real astronomical data.  Bayesian
methods have historically been limited by lack of computing power; this is demonstrably no longer the case,
giving us an array of new statistical tool with which to improve
astronomical source extraction and hence the astrophysical science that
depends upon it.

The methods described in this paper have been implemented as a
beta-version, publicly available software tool (written in IDL).  The
code plus associated documentation and test data can be obtained from
the following URL:

http://astronomy.sussex.ac.uk/$\sim$rss23/sourceMiner\_v0.1.2.0.tar.gz

\acknowledgements
Richard Savage thanks the Particle Physics and Astronomy Research
Council (PPARC) for support under grant PPA/G/S/2002/00481.
Seb Oliver thanks the Leverhulme Trust for support in the form of a
Leverhulme research fellowship.
We thank Philippe Andre, Bruce Sibthorpe and Tim Waskett for their kind
provision of the simulated flux image shown in Figure
\ref{fig:sourceDetection}.
We thank Mike Hobson, Emmanuel Bertin and Douglas Scott for useful
comments and discussions.
We thank the anonymous referee for useful comments.

\label{lastpage}
\bibliography{../../bibtex_files/inference_refs,../../bibtex_files/sourceExtractionRefs,../../bibtex_files/IR_refs}
\bibliographystyle{../mn2e}
\end{document}